\documentstyle[eqsecnum,preprint,aps]{revtex}

\def\d3q#1{\frac{d^3 q_#1}{(2\pi)^3}}
\def\lesquig{  {\ \lower-1.2pt\vbox{\hbox{\rlap{$<$}\lower5pt\vbox{
\hbox{$\sim$}}}}\ } }

\def\footnoterule{\kern-3pt \hrule width \hsize \kern6.2pt}

\def\pmb#1{\setbox0=\hbox{$#1$}%
\kern-.025em\copy0\kern-\wd0
\kern.05em\copy0\kern-\wd0
\kern-.025em\raise.0433em\box0 }
\tighten

\begin{document}

\title{Fermion Production in the Background of \\
Minkowski Space  Classical Solutions in\\
Spontaneously Broken Gauge Theory}

\author{Edward Farhi
and Jeffrey Goldstone\footnote[1]{farhi@mitlns.mit.edu
and goldstone@mitlns.mit.edu.~~
This work is supported in part by funds provided by the
U.S.~Department of Energy (D.O.E.)
under cooperative agreement \#DF-FC02-94ER40818.}}
\smallskip
\address{Center for Theoretical Physics\\
Laboratory for Nuclear Science and Department of Physics\\
Massachusetts Institute of Technology\\
Cambridge, MA\ \ 02139}

\author{Sam Gutmann}
\smallskip
\address{Department of Mathematics\\
Northeastern University\\
Boston, MA\ \ 02115}

\author{Krishna Rajagopal\footnote[2]{rajagopal@huhepl.harvard.edu.~~
Junior Fellow, Harvard Society of Fellows.~~Research supported in part
by the Milton Fund of Harvard University and by
the National Science Foundation under grant PHY-92-18167.}}
\smallskip
\address{Lyman Laboratory of Physics\\
Harvard University\\
Cambridge, MA\ \ 02138}

\author{Robert Singleton, Jr.\footnote[3]{bobs@cthulu.bu.edu.~~Research
supported in
part by the D.O.E. under contract \#DE-FG02-91ER40676 and
by the Texas National Research Laboratory Commission
under grant RGFY93-278.}}
\smallskip
\address{Department of Physics\\
Boston University\\
Boston, MA\ \ 02215}

\maketitle

\setcounter{page}{0}
\thispagestyle{empty}

\vfill

\noindent CTP\#2370

\noindent HUTP-94/A038 \hfill Submitted to {\it Physical Review} {\bf D}

\noindent BU-HEP-94-30 \hfill Typeset in REV\TeX
\eject

\vfill

\begin{abstract}

We investigate fermion production in the background of
Minkowski space solutions to the equations of motion of $SU(2)$
gauge theory spontaneously broken via the Higgs mechanism.
First, we attempt to evaluate the topological charge $Q$
of the solutions.
We find that for solutions
$Q$ is not well-defined
as an integral over all space-time.
Solutions can
profitably be characterized by the (integer-valued) change
in Higgs winding number $\Delta N_H$.  We show that
solutions which dissipate at early and late times and
which have nonzero $\Delta N_H$ must have at least
the sphaleron energy.  We show that if we couple a
quantized massive chiral fermion to a classical background given by
a solution, the number of fermions produced is $\Delta N_H$,
and is not related  to $Q$.

\end{abstract}

\vfill

\eject

\baselineskip 24pt plus 2pt minus 2pt

\section{Introduction}
\label{SectionIntro}

In this paper we study general properties of classical
solutions in $SU(2)$ gauge theory with spontaneous symmetry
breaking introduced via the Higgs mechanism.
The bosonic sector of the model we consider is
that of the standard electroweak
theory without the $U(1)$ gauge boson.
We work entirely in Minkowski space and look
at solutions with finite energy.
We add a massive quantized
$SU(2)$ doublet chiral fermion, and discuss
fermion production in
the background of classical
gauge and Higgs field solutions.
We find that the change in fermion
number is not determined by the topological
charge, but rather equals the change in the
winding of the Higgs field.

In previous work \cite{FKRS},
classical solutions in
$SU(2)$ gauge theory with no Higgs field were studied.
These solutions have
the property that in the far past and far future they can be
described as spherical shells which propagate without distortion.
Furthermore, such solutions have nonzero,
non-integer topological charge.  In this paper
the Higgs field is included and we
do not restrict ourselves to  the spherical
ansatz.  As we will see, the
solutions we consider here are qualitatively
different than those considered in Ref. \cite{FKRS}.

The topological charge is defined as
  \begin{equation}
  \label{IntroQ}
   Q = \frac{g^2}{32\pi^2}
  \int d^4 x \,
  \epsilon^{\mu\nu\alpha\beta} \,
  {\rm Tr}(F_{\mu\nu} F_{\alpha\beta}) \ .
  \end{equation}
The usual argument \cite{J} that leads to integer
topological charge requires that the region
in space-time where the energy density is nonzero
be bounded.  For solutions to the equations
of motion, energy is conserved and the energy
computed on {\it any} equal time surface is non-vanishing.
Thus, for Minkowski space solutions, we have no reason
to expect integer values of $Q$.

The integral in (\ref{IntroQ}) is over all of
space-time.  We can attempt to evaluate
it as the limit of
a sequence of integrals taken over larger and larger
regions of space-time.
$Q$ is well-defined if
we get the same finite result for (\ref{IntroQ})
no matter what
sequence of integration regions is chosen
so long as in the limit
all of space-time is
included.
We find that $Q$ evaluated on
solutions to the classical equations
of motion for spontaneously broken $SU(2)$
gauge theory is not well-defined in this sense.
Specifically, we first evaluate (\ref{IntroQ}) by
doing the integral inside a rectangular space-time box whose
size we take to infinity, and
obtain a finite result.  Then, we redo the calculation
using a differently shaped convex box
and obtain
a {\it different} finite result.
Therefore, the topological charge of solutions
to the equations of motion cannot be defined.

Suppose we couple a quantized chiral fermion
to the classical gauge and Higgs field backgrounds
considered in this paper.  The anomaly equation
relates $Q$ to $\int d^4x\, \partial_\mu J^\mu (x)$
for an appropriately defined fermion current.
However, for solutions to the equations of motion
$Q$ is not defined and the implications
of the anomaly equation for fermion production are
murky.
Consider a continuous sequence of
configurations (which is not a solution)
with the gauge field chosen so that
the background has a well-defined nonzero $Q$
and
with the Higgs field a nonzero constant.
Because the fermion mass is generated by
a Yukawa coupling to the Higgs field, the
fermion mass is also a nonzero constant.
If we make the fermion mass large while
keeping the background gauge field fixed,
surely no fermionic level will cross the mass
gap as we follow the sequence from beginning to
end.
Thus, in a theory with a massive fermion
in a background that does not go from
vacuum to vacuum, fermion
production cannot simply depend on $Q$,
which is determined only by the gauge field.
Since the Yukawa coupling is
proportional to the fermion mass, it seems reasonable
that the background Higgs field plays a crucial
role in fermion level-crossing\cite{guralnik}.
In fact, we will argue in Section 5 that the number
of fermions produced in a background given by a solution
which dissipates at early and late times is
equal to the change in the winding
number of the Higgs field.

Solutions to the equations of motion are
particular examples of continuous sequences
of configurations parameterized by $t$ which
move through
a configuration space described by
the gauge field $A_\mu({\bf x})$ and the
Higgs field $\Phi({\bf x})$.
Vacuum configurations
can be characterized by the integer-valued
winding number of the gauge field.
Sequences of configurations beginning
and ending in distinct
vacua have a
nonzero, integer-valued
topological charge given by the difference between
the winding number of the gauge field in
the final and initial vacuum configurations.
The sphaleron \cite{manton}
is the lowest energy point on the barrier
which a path in configuration
space connecting two topologically distinct vacua
must surmount.
Therefore, vacuum to vacuum sequences of configurations
which have $Q\neq 0$ must pass over the sphaleron barrier.
However, $Q$ cannot profitably be used to
characterize solutions since for solutions
$Q$ is not well-defined.

The integer-valued Higgs winding
number can be used to
characterize even nonvacuum configurations.
Let $\varphi = (\varphi_1,\varphi_2)$
be the usual Higgs doublet and define
\begin{equation}
\label{Phidef}
\Phi({\bf x}) \equiv \left( \matrix{ \varphi_2^* & \varphi_1 \cr
  -\varphi_1^* & \varphi_2 } \right) \ .
\end{equation}
At
points ${\bf x}$ where $\rho({\bf x})\equiv (\varphi_1^*\varphi_1 +
\varphi_2^*\varphi_2)^{1/2}$ is non-vanishing, the Higgs
field can be associated with a special unitary matrix
$U({\bf x})$ through
  \begin{equation}
  \label{Udef}
  \Phi({\bf x})
= \rho({\bf x}) \, U({\bf x}) \ .
  \end{equation}
We consider only those configurations for
which the fields approach their vacuum values in the
$|{\bf x}|\rightarrow\infty$ limit.
Without loss of generality, we can impose the
boundary conditions
  \begin{mathletters}
  \label{UAbc}
  \begin{eqnarray}
  \lim_{|{\bf x}|\rightarrow\infty}U({\bf x}) &=&  1 \ ,
  \label{Ubc} \\
  \lim_{|{\bf x}|\rightarrow\infty}A_\mu({\bf x}) &=&  0 \ ,
  \label{Abc}
  \end{eqnarray}
\end{mathletters}
and accordingly only consider gauge transformations which
approach unity as $|{\bf x}|\rightarrow\infty$.
Configurations with
$\rho\neq 0$ throughout
space can be characterized
by the Higgs winding number
  \begin{equation}
  \label{NHdef}
  N_H = w[U]
\end{equation}
where the integer-valued
winding number of a special unitary matrix $U$ satisfying
(\ref{Ubc}) is
\begin{equation}
\label{windingdef}
w[U]
=\frac{1}{24\pi^2}\int {\rm d}^3x  \, \epsilon^{ijk}\,
  {\rm Tr}\biggl( U^\dagger \partial_i U \,
  U^\dagger \partial_j U \,U^\dagger \partial_k U \biggr) \ .
  \end{equation}
$N_H$ is gauge invariant under small gauge transformations,
but  it changes under large gauge transformations.

Now, we return to our discussion of
solutions to the equations of motion,
which can be viewed as
continuous sequences of configurations
parameterized by $t$.
Solutions typically
have zeros of $\rho$ isolated in space-time. At times
when $\rho$ is everywhere nonvanishing
we define $N_H(t)$ as the Higgs
winding number of the configuration
at time $t$.
Although solutions with nonzero energy
do not begin and end in vacua,
those which we consider have
their energy density approach zero uniformly
in space in the $t\rightarrow\pm\infty$
limits.  We refer to this behavior as dissipation,
noting that it is the energy density which
dissipates while the energy is conserved
and may be large (relative to the sphaleron energy.)
In this sense, the solutions we consider
begin and end {\it near} vacua.
At very early and late times
$\rho$ is approaching its vacuum value and
is everywhere nonzero.  Therefore,
$N_H(t)$ becomes
constant in time in the far past and in the far future
and we can define
  \begin{equation}
  \Delta N_H = \lim_{t\rightarrow\infty}N_H(t) -
  \lim_{t\rightarrow -\infty}N_H(t) \ .
  \label{DeltaNHdef}
  \end{equation}
Note that this difference is gauge invariant
even under large gauge transformations.
We will show in Section 4 that if a
solution which dissipates at early and late times
has $\Delta N_H\neq 0$ then it must
begin and end near distinct vacua and must therefore
have
at least the sphaleron energy.

What do we learn in this paper about
fermion production in the background of
solutions to the Minkowski space
equations of motion which
dissipate at early and late times?  Such
solutions do not have well-defined
topological charge; they do have a well-defined
change in Higgs winding; it is $\Delta N_H$ which
counts the number of fermions produced.
For a solution to have nonzero
$\Delta N_H$, it must have at least
the sphaleron energy.  This means
that nonzero $\Delta N_H$ and the
associated fermion production
cannot occur at finite order in perturbation
theory.

The logical next step in our investigation
is to consider quantum scattering of massive
gauge bosons, rather than the
classical field configurations which are
the subject of most of this paper.
Because we have introduced the Higgs field into
the theory, the asymptotic states in the quantum
theory are in fact gauge bosons (and not glueballs)
and we can therefore begin to consider quantum
scattering.  We show that the unexpected
behavior of the topological charge
of classical solutions (namely that it is not
defined) has an analogue in quantum scattering.
We attempt to
define a quantum operator
for which the difference between
its expectation value in the initial
and final states in some scattering process
measures the topological charge associated
with that process and find that we cannot do so
in a Lorentz invariant fashion.

\vskip .3in

\section{Solutions to the Equations of Motion}
\label{eqsection}

In this section, we discuss the behavior of solutions
to the classical equations of motion for $SU(2)$ gauge
theory spontaneously broken via the Higgs mechanism.
The action is
  \begin{equation}
  \label{Action}
  S=\int d^4x\,  \left\{ -\frac{1}{2} {\rm Tr}(F^{\mu\nu}F_{\mu\nu})  +
  \frac{1}{2}{\rm Tr}\,\left(D^\mu\Phi\right)^\dagger  D_\mu\Phi
  - \frac{\lambda}{4} \left( {\rm Tr} \Phi^\dagger \Phi  -
  v^2 \right)^2 \right\} \ ,
  \end{equation}
where the $2\times 2$ matrix $\Phi$ is related to
the Higgs doublet $\varphi$ by equation (\ref{Udef})
and where
  \begin{eqnarray}
  \label{Fmunu}
  F_{\mu\nu} &=& \partial_\mu A_\nu - \partial_\nu A_\mu - i g [A_\mu,A_\nu]
  \nonumber\\
  D_\mu \Phi &=& (\partial_\mu - i g A_\mu)\Phi
  \end{eqnarray}
with $A_\mu = A_\mu^a \sigma^a/2$. We use the conventional
metric $\eta_{\mu\nu}={\rm diag}(1,-1,-1,-1)$.
The action is invariant under the
transformation
\begin{eqnarray}
A_\mu &\rightarrow & G A_\mu G^\dagger + \frac{i}{g}
G \partial_\mu G^\dagger \equiv A_\mu^G \nonumber \\
\Phi &\rightarrow & G \Phi
\label{gaugetrans}
\end{eqnarray}
where $G(x) \in SU(2)$.
The equations of
motion are
  \begin{mathletters}
  \begin{eqnarray}
  D_\mu F^{\mu\nu} &=& \frac{i g}{4}\left[  \Phi \left(D^\nu \Phi
\right)^\dagger
  - \left( D^\nu \Phi \right) \Phi^\dagger
  \right] \ , \\
  D_\mu D^\mu \Phi &=& -\lambda \left({\rm Tr}\,\Phi^\dagger
  \Phi - v^2 \right) \Phi \ ,
  \end{eqnarray}
  \label{eqsofmotion}
  \end{mathletters}
where
  \begin{equation}
\label{DmuFmunu}
  D_\mu F^{\mu \nu} = \partial_\mu F^{\mu \nu} -
  i g [A_\mu,F^{\mu \nu}] .
  \end{equation}

We expect a ``typical'' solution to dissipate both in
the far future and in the far past. By dissipation we
mean that at early and late times
the energy density approaches zero
uniformly throughout space.
Not all solutions exhibit
dissipation.  For example, the sphaleron is a static solution
and therefore its energy density is constant in time.
One can also imagine solutions which are asymptotically
equal to the sphaleron for early (late) times but which
dissipate at late (early) times.
Thus, by restricting ourselves to solutions
which dissipate both in the future and the
past we are excluding
some solutions from our treatment.

For the solutions we wish to treat,
at early and late times the magnitude of the Higgs field is
everywhere close to its vacuum value and, in particular,
does not vanish.
This suggests that we work in unitary gauge, in which
$U$  defined in (\ref{Udef})  is set to unity.
However, if $\Delta N_H \neq 0$, then it is impossible to
choose a gauge
in which $U=1$ both in the future and in the past.
It is, however, possible to choose one gauge in which $U=1$
in the far past and another gauge in which $U=1$ in the
far future.
These two gauges will differ by a large gauge transformation
with winding $\Delta N_H$.

At either early or late times we can go into the
unitary gauge, in which $U=1$ and the
equations of motion for $A_\mu$  and $\rho$ are
\begin{mathletters}
  \label{unitaryeqns}
  \begin{equation}
  D_\mu F^{\mu \nu} + \frac{g^2}{2}\rho^2  A^\nu = 0 \ ,
  \label{EqUnitarya}
  \end{equation}
  \begin{equation}
  \partial_\mu \partial^\mu \rho
  - \frac{1}{2} g^2 \rho \,{\rm Tr}\left( A^\mu A_\mu \right)
  +  2 \lambda  \left( \rho^2 - \frac{v^2}{2}
  \right) \rho = 0 \ .
  \label{EqUnitaryb}
  \end{equation}
  \end{mathletters}
There is only one vacuum
configuration in unitary gauge: $A_\mu = 0$, $\rho = v/\sqrt{2}$.
We therefore expand the equations of
motion as power series in $A_\mu$ and
in the shifted field
\begin{equation}
h \equiv \rho - \frac{v}{\sqrt{2}} \ .
\label{hdef}
\end{equation}
To linear order, the equations of motion are
\begin{mathletters}
\label{lineqns}
\begin{eqnarray}
\partial^\mu \left( \partial_\mu A_\nu^{\rm lin} -
\partial_\nu A_\mu^{\rm lin} \right) + m^2 A_\nu^{\rm lin}
&=& 0 \ , \label{linAeqn} \\
\left(\partial^\mu \partial_\mu + m_h^2 \right) h^{\rm lin} &=& 0 \ ,
\label{linheqn}
\end{eqnarray}
\end{mathletters}
where
\begin{equation}
m = \frac{1}{2}\,gv ~~~~~~~~m_h = \sqrt{2 \lambda}\,v \ .
\label{masses}
\end{equation}
By taking the divergence
of (\ref{linAeqn}), we get
  \begin{equation}
   \label{lorentzgauge}
  \partial^\mu A_\mu^{\rm lin} = 0 \ .
  \end{equation}
Equation (\ref{lorentzgauge}) is the same equation as
the Lorentz gauge condition, which here arises
as a consequence of the linearized equations of motion in
unitary gauge.  Using (\ref{lorentzgauge}),
equation (\ref{linAeqn}) becomes
\begin{equation}
\left(\partial_\nu\partial^\nu + m^2\right)A_\mu^{\rm lin} = 0 \ .
\label{eqsimple}
\end{equation}
Note that (\ref{eqsimple}) are the
equations of motion of independent massive vector fields
labeled by an $SU(2)$ index  but there is no
remaining gauge invariance.

The solution to (\ref{eqsimple}) takes
the form
  \begin{equation}
  \label{zeroth}
  A_\mu^{\rm lin}(x) = \frac{1}{(2\pi)^3}\int \frac{d^3 k}{2\omega_k} \,
  \left[ \, e^{-i k \cdot x} \epsilon_\mu(k)+ e^{i k \cdot x}
  \epsilon_{\mu}^*(k) \, \right] \ ,
  \end{equation}
where  $k^\mu=(\omega_k,{\bf k})$ with $\omega_k = ({\bf k}^2 +
m^2)^{1/2}$, and where the $SU(2)$ valued polarization vector
$\epsilon_\mu$ transforms as a vector under Lorentz
transformations and  satisfies $\epsilon_\mu k^\mu=0$.
There is a large class of solutions to (\ref{eqsimple})
for which
$A_\mu^{\rm lin}({\bf x},t)\rightarrow 0$
uniformly in ${\bf x}$ as
$t\rightarrow\pm\infty$.
This will certainly be the case if at
some time $t_0$, $A_\mu^{\rm lin}({\bf x},t_0)$
and $\partial_t A_\mu^{\rm lin}({\bf x},t_0)$
are sufficiently smooth and vanish for $|{\bf x}|$
greater than some $R$.
Under analogous conditions,
solutions $h^{\rm lin}$ to (\ref{linheqn}) also
approach zero uniformly in ${\bf x}$ at early and
late times.  Also,
the energy density of
these solutions to the linearized equations
vanishes uniformly in ${\bf x}$
as $t\rightarrow\pm\infty$.

We now have a picture of the behavior of
solutions to the equations of motion which dissipate.
We choose to work in a gauge in which $U=1$
in the far past.  In this gauge,
$A_\mu$ is well approximated at early times
by (\ref{zeroth}) with some polarization tensor
$\epsilon_\mu^p(k)$.  This ``past'' polarization
tensor may be such that as the solution
evolves forward in time, energy
that was widely separated in space comes together,
energy densities grow, and nonlinear effects
become important.  If the nonlinearities conspire
to prevent the energy density from dissipating
in the far future, as for example if a sphaleron
is created, then we exclude the solution from
our discussion.  It is more likely, however,
that at late times the energy density is
once again spread over a large region
of space.
There is then a gauge in which
$A_\mu$ is
again well approximated by (\ref{zeroth}) this time with
a different polarization  tensor $\epsilon_\mu^f(k)$.
In the gauge in which $U=1$ in the far past, however,
$A_\mu$ in the far future is given by
\begin{equation}
  A^f_\mu(x) = \frac{1}{(2\pi)^3}\int \frac{d^3 k}{2\omega_k} \,
  U_f(x)\left[ \, e^{-i k \cdot x} \epsilon^f_\mu(k)+ e^{i k \cdot x}
  \epsilon^{f *}_{\mu}(k) \, \right] U_f^\dagger(x)
  \, + \, \frac{i}{g} U_f(x) \partial_\mu U_f^\dagger(x)  \ ,
\label{Afuture}
\end{equation}
where $U_f(x)$ is an $SU(2)$ valued gauge function
satisfying the boundary condition (\ref{Ubc})
and where $w[U_f]=\Delta N_H$.

It is possible to go much farther in solving the
equations of motion.  They can be expanded
order by order in $g$ and $\lambda$, and
the resulting equations can be solved
for specified initial data
using Greens function methods.  This procedure
can be organized using tree-level Feynman diagrams.
Given $\epsilon_\mu^p(k)$
(and suitable initial
data for $h$) one could solve order by order
in $g$ and $\lambda$ for $\epsilon_\mu^f(k)$.
In discussing topological properties of solutions
in the rest of this paper we will not need to
obtain $\epsilon_\mu^f(k)$ explicitly for a given
$\epsilon_\mu^p(k)$, and therefore
we will not need to solve the
higher order equations of motion.

\vskip .3in

\section{Undefined Topological Charge}
\label{pertsection}

The topological charge of a sequence of configurations
parameterized by $t$ is defined by
  \begin{equation}
  \label{topQ}
  Q = \frac{g^2}{32\pi^2}\int d^4 x \,
  \epsilon^{\mu\nu\alpha\beta}
  {\rm Tr}(F_{\mu\nu} F_{\alpha\beta}) \ ,
  \end{equation}
where the integration is over all space-time.
The integrand in (\ref{topQ}) is a
total divergence and $Q$ can be written
  \begin{equation}
  \label{Qd4Kmu}
  Q =\int d^4 x \,
 \partial_\mu K^\mu \ ,
  \end{equation}
where the topological current is
  \begin{equation}
  K^\mu =  \frac{g^2}{32\pi^2} \, \epsilon^{\mu\nu\alpha\beta} \,
  {\rm Tr}( A_\nu \partial_\alpha A_\beta + \frac{2}{3}\, ig\,
  A_\nu A_\alpha A_\beta) \ .
  \end{equation}
For solutions to the equations of motion
described in the previous section,
we now
perform the integration in (\ref{topQ}) in
two different ways and obtain
finite, but different, answers.

In our first attempt at evaluating (\ref{topQ}),
we  do the integral in a finite space-time
box extending from $t=-T$ to $t=T$, then
take the spatial extent of the box to infinity,
and finally take $T$ to infinity.
We call the result of this calculation $Q_1$.
We consider only solutions with compact support,
by which we mean that at any time $t$
the fields have vacuum
values for $|{\bf x}|$ greater than some ($t$ dependent) $R$.
We work in a gauge such that for $|{\bf x}|>R$,
$A_\mu =0$ (and $U=1$) and
therefore there is no contribution
to (\ref{Qd4Kmu}) from the surface at spatial
infinity. It is convenient to define the Chern-Simons number
\begin{equation}
N_{CS}(t)= \int d^3x\, K^0({\bf x},t) \ ,
\label{NCSdef}
\end{equation}
keeping in mind that $N_{CS}$ is not
integer-valued. Then,
\begin{equation}
Q_1 = N_{CS}^f - N_{CS}^p \ ,
\label{QdiffW}
\end{equation}
where
\begin{equation}
\label{NCSfdef}
N_{CS}^f = \lim_{T\rightarrow\infty}N_{CS}(T)
\end{equation}
and
\begin{equation}
\label{NCSpdef}
N_{CS}^p = \lim_{T\rightarrow -\infty}N_{CS}(T) \ .
\end{equation}

We are considering solutions that have
the property that $A_\mu({\bf x},t)
\to 0$ for any fixed ${\bf x}$ as $t \rightarrow -\infty$, and one might
naively conclude that $N_{CS}^p$ is zero. However,
care is needed when interchanging time limits and spatial
integrals --- equation (\ref{NCSpdef})
requires us to do the $d^{\,3}x$ integral
at finite time, and only then to take the
$T\rightarrow -\infty$ limit.
For example, the energy
density also has the property that it vanishes at any fixed spatial
point in the infinite past; however, its spatial integral
is constant in time and nonzero.

At early
times the solutions we are discussing
are well approximated by solutions to the linear
equations.
We choose the gauge such that in the far past
$A_\mu$ is of the form (\ref{zeroth}) with
polarization vector $\epsilon^p$.
In evaluating the right side of
(\ref{NCSpdef}),
we find that only terms quadratic in
$A_\mu$ which involve products of $\epsilon^p$ and $\epsilon^{p*}$
are nonzero.  All other terms, including the cubic terms
in $A_\mu$, vanish by the Riemann-Lebesgue theorem.
We obtain
\begin{mathletters}
\begin{equation}
N_{CS}^p =
  \frac{g^2}{32\pi^2} \frac{1}{(2\pi)^3}\int \frac{d^3 k}{2\omega_k}  \,
  n_\mu\epsilon^{\mu\nu\alpha\beta}
  \,\frac{ik_\nu}{\omega_k}\, {\rm Tr}
  \left[ \epsilon_\alpha^{p*}(k)
  \epsilon_\beta^{p}(k) \right] \ ,
\label{NCSp}
\end{equation}
where $n^\mu \equiv (1,0,0,0)$.
The trace is over the $SU(2)$ indices
on the polarization vectors.
To evaluate $N_{CS}^f$, we use (\ref{Afuture}) in (\ref{NCSfdef}),
noting that gauge transforming a configuration by a
gauge function $U$ adds $w[U]$ to the Chern-Simons number,
and obtain
\begin{equation}
N_{CS}^f  =
  \frac{g^2}{32\pi^2} \frac{1}{(2\pi)^3}\int \frac{d^3 k}{2\omega_k}  \,
  n_\mu\epsilon^{\mu\nu\alpha\beta}
  \,\frac{ik_\nu}{\omega_k}\,
{\rm Tr}
  \left[ \epsilon_\alpha^{f*}(k)
  \epsilon_\beta^{f}(k) \right]  \,+\, w[U_f] \ ,
\label{NCSf} \\
\end{equation}
\end{mathletters}
where the polarization vectors
$\epsilon_\mu^f(k)$ characterize the solution in the
far future.
The topological charge given by (\ref{QdiffW}) is generically
nonzero, finite, and not an integer, and one may be tempted
to stop here.

However, we now redo the calculation using a different sequence
of integration regions.  Whereas above we considered
space-time boxes bounded in time by $t=\pm T$,
we now consider
space-time regions bounded by $t'=-T$ and $t=T$, where $t'$ is a
Lorentz transform of $t$.  For concreteness, we take the Lorentz
transformation to be a boost of velocity $\beta$ in
the 3-direction, and so have $t'=\gamma(t + \beta x^3 )$
with $\gamma=1/\sqrt{1-\beta^2}$.
The surfaces $t=T$ and $t'=-T$ intersect and
we take the integration region to be the wedge which includes
the origin ${\bf x}=0$, $t=0$.  Note that as $T$ goes to infinity,
this wedge includes all of space-time, and we call the result
of this calculation $Q_2$.
As before, we do the integral
as a surface integral.  For solutions with compact support,
there is no contribution from the surface at spatial infinity.
Also, note that for sufficiently large $T$ the surfaces
$t=T$ and $t'=-T$ intersect in a region of space-time
where $A_\mu=0$, and the parts of these surfaces which
do not bound the wedge over which we are integrating have
$A_\mu=0$.
Therefore, doing the integral in (\ref{topQ}) over the
wedge between $t=T$ and $t'=-T$ which includes the origin
and then taking the $T\rightarrow\infty$
limit yields
\begin{equation}
Q_2 = {N}_{CS}^f - {N'}_{CS}^p
\label{Qprime}
\end{equation}
with ${N}_{CS}^f$ as before and with
\begin{equation}
\label{newNCSprime}
{N'}_{CS}^p = \lim_{T\rightarrow\infty} \int d^3x' {K'}^0({\bf x}',-T)
\end{equation}
where ${x'}^\mu$ is the Lorentz transform of $x^\mu$ and
where ${K'}^\mu$ is the Lorentz transform of $K^\mu$.
For (\ref{newNCSprime}) we obtain an expression
identical in form to (\ref{NCSp})
but with $k_\mu$ and $\epsilon_\mu$
replaced by their Lorentz transforms $k'_\mu$ and
$\epsilon'_\mu$.
Thus,
  \begin{equation}
  \label{NCSinfprime}
  {N'}_{CS}^p = \frac{g^2}{32\pi^2} \frac{1}{(2\pi)^3}\int \frac{d^3 k'}
  {2 \omega_{k'}}  \,
  n_\mu
  \epsilon^{\mu\nu\alpha\beta}
  \,\frac{ik'_\nu}{\omega_{k'}}\, \,{\rm Tr}
  \left[ {\epsilon '}^{p*}_\alpha(k')
  {\epsilon '}^{p}_\beta(k') \right]
  \end{equation}
(Note that in (\ref{NCSinfprime}) we mean $n_\mu$ and not
$n'_\mu$.)
Using the fact that $k_\mu$ and $\epsilon_\mu$
are Lorentz vectors and that $d^3k/\omega_k$
is Lorentz invariant we write
\begin{equation}
\label{Qprime2}
  {N'}_{CS}^p = \frac{g^2}{32\pi^2} \frac{1}{(2\pi)^3}\int \frac{d^3 k}
  {2 \omega_{k}}  \,
  \tilde n_\mu
  \epsilon^{\mu\nu\alpha\beta}
\,  \frac{ik_\nu}{\omega_{k'}}
\,{\rm Tr} \left[ \epsilon^{p*}_\alpha(k)
  \epsilon^p_\beta(k) \right] \ ,
  \end{equation}
where $\omega_{k'} =\gamma(\omega_k + \beta k^3)$
and $\tilde n^\mu \equiv
(\gamma,0,0,-\gamma\beta)$.
Then, using $\epsilon^p_\mu k^\mu =0$
to eliminate $\epsilon^p_0$,
we find
\begin{equation}
\label{NCSinflorentz}
  {N'}_{CS}^p = N_{CS}^p +
\frac{g^2}{32\pi^2} \frac{1}{(2\pi)^3}\int \frac{d^3 k}{2\omega_k}
	\, \frac{i\beta m^2}{\omega_k
	(\omega_k + \beta k^3) }
         {\rm Tr}\, \left[
	\epsilon^p_1(k) \epsilon^{p *}_2(k) - \epsilon^p_2(k)
	\epsilon^{p *}_1(k)
	\right] \ ,
  \end{equation}
where $N_{CS}^p$ is the previous result (\ref{NCSp}).
We have found ${N'}_{CS}^p\neq N_{CS}^p$,
and therefore $Q_2 \neq Q_1$.

We have now evaluated the integral (\ref{topQ})
using two different sequences of regions of integration.
Once the $T\rightarrow\infty$ limit has been taken,
$Q_1$ and $Q_2$ are integrals of the same integrand
over all space-time, and yet they differ.
This implies that for solutions with compact support
$\int d^4 x | \epsilon^{\mu\nu\alpha\beta}
{\rm Tr}(F_{\mu\nu}F_{\alpha\beta}) |$
is infinite, {\it i.e.} the integral (\ref{topQ}) is not absolutely
convergent.
Some integrals which fail to be absolutely convergent
can be defined
unambiguously by integrating over larger and larger
convex regions which grow to encompass all of
space-time.  (As an example,
consider $\int dt\,dx \exp (-x^2) (\sin t) / t$.\,)
Since $Q_1\neq Q_2$, however, the integral (\ref{topQ})
cannot be defined in this way and thus fails to
be absolutely convergent in a particularly bad way.

The reader may
wonder whether the topological charge could be
defined by requiring that the integral (\ref{topQ}) be
evaluated between parallel space-like surfaces as
for $Q_1$.
To investigate this, evaluate the integral between
$t'=-T$ and $t'=T$, take the $T\rightarrow \infty$ limit,
and call the result $Q_3$.  $Q_3$ is given by the difference
between   ${N'}_{CS}^p$  of (\ref{NCSinflorentz}) and
the analogous ${N'}_{CS}^f$.
Is it possible for $Q_3$ and $Q_1$ to be equal
for arbitrary Lorentz transformations?
For this to occur,
it would be necessary that the integral on the
right hand side of (\ref{NCSinflorentz}) be the
same if $\epsilon^p$ is replaced by $\epsilon^f$ for all
values of $\beta$.  Although $\epsilon^f$ is determined
by $\epsilon^p$ so they cannot be viewed as independent,
we know of no conservation law strong enough to
guarantee this for all $\beta$.  Thus we believe that
$Q_3\neq Q_1$.

In summary, the integral (\ref{topQ}) cannot be defined
as an absolutely convergent integral over all space-time
and cannot be defined as the limit of the integral
over arbitrary sequences
of larger and larger convex regions of space-time.
We know of no reasonable way to define the topological
charge of solutions to the equations of motion.

Note, however, that by setting $m=0$ in equation (\ref{NCSinflorentz})
we see that in
the massless theory, $Q_1$ and $Q_2$ {\it are}
identical.
In fact, it can be shown that in the massless case
$\int d^4 x | \epsilon^{\mu\nu\alpha\beta}
{\rm Tr}(F_{\mu\nu}F_{\alpha\beta}) |$
is finite and therefore
the topological charge
is well-defined.  This result can be established by
using the explicit form of the massless Green's function to show
that for solutions with compact support,
$\int d^3 x | \epsilon^{\mu\nu\alpha\beta}
{\rm Tr}(F_{\mu\nu}F_{\alpha\beta}) |$
falls like $1/|t|^2$ for large $|t|$.
Since the topological charge of solutions in the massless
theory is well-defined, the explicit calculations done
in Ref. \cite{FKRS} are unambiguous.

Note that equation (\ref{NCSinflorentz}) can also be seen
as a demonstration that the early time Chern-Simons
number is not Lorentz invariant in the massive theory,
but is Lorentz invariant in the massless theory.
(The quantum analogue of this fact is discussed in Section 6.)
Given a Lorentz vector $K^\mu$, the standard
proof that $\int d^{\,3} x\, K^0(x)$
is Lorentz invariant requires that $K^\mu$
be conserved, {\it i.e.} that $\partial_\mu K^\mu = 0$.
In this sense, it is
actually the Lorentz invariance of $N_{CS}(t)$
at early and late times
in the massless theory which is more surprising
than the lack of Lorentz invariance of $N_{CS}(t)$
at early and late times in the massive theory,
since $K^\mu$ is not a conserved current in either
case.

\vskip .3in

\section{${\pmb\Delta}{\bf N}_{\bf H}$
and the Sphaleron}
\label{sphalsection}

The sphaleron was introduced \cite{manton} in the context
of studying sequences of configurations in spontaneously
broken $SU(2)$ gauge theory.  One way of stating
the result of Ref. \cite{manton} is that if
a continuous
sequence of configurations which begins and ends
in vacuum has nonzero topological charge, then
it must include configurations with at least the
sphaleron energy.\footnote{
 Any sequence of
 configurations which begins and ends in distinct
 vacua has some maximum energy.
 We assume, as is generally assumed, that this
 maximum energy cannot be arbitrarily close to zero.
 It is also generally believed that
 of all the maximum energy
 configurations, the one with the lowest energy is the
 sphaleron of Ref. \cite{manton}
 if $\lambda/g^2$ is sufficiently small, while for
 larger $\lambda/g^2$ the deformed sphalerons
 discovered by Yaffe \cite{yaffe} play this role.
 What we refer to as the sphaleron is the highest
 energy configuration on the path of minimum
 maximum energy,
 independent of the form of this configuration.}
However, topological charge cannot be used to
distinguish solutions to the equations of motion which
pass over the sphaleron barrier
from those which do not, because
solutions do not have well-defined
topological charge.
As we discussed in the Introduction,
solutions can be characterized by the
gauge invariant integer $\Delta N_H$,
and we show in this section that for
solutions which dissipate at early and
late times $\Delta N_H \neq 0$ only for
solutions which cross the sphaleron
barrier and consequently have
energies exceeding that of the sphaleron.

Consider configurations specified by
$A_i({\bf x})$ and $\Phi({\bf x})$
where we always take $A_0=0$.
Following Ref. \cite{manton},
we define a non-negative
potential energy functional $V[A_i,\Phi]$
by taking the Hamiltonian associated with (\ref{Action}),
setting $A_0=0$
and dropping all terms involving time derivatives
of fields. (Note that configurations are specified by
the values of the fields, without reference
to their time derivatives.)
We obtain
\begin{equation}
V[A_i,\Phi]=\int d^3x \left\{
\frac{1}{2}{\rm Tr}\,\left(F_{ij}F_{ij}\right)
+ \frac{1}{2}{\rm Tr} \left(D_i \Phi \right)^\dagger D_i \Phi
+ \frac{\lambda}{4} \left( {\rm Tr}\Phi^\dagger\Phi - v^2 \right)^2
\right\} \ ,
\label{Veqn}
\end{equation}
which we refer to as the energy of the configuration
$A_i$, $\Phi$.

A continuous  sequence of configurations
parametrized by $t$ with
$N_H(t_1)\neq N_H(t_2)$
can consist entirely
of configurations with arbitrarily low energy
for $t$ between $t_1$ and $t_2$,
as we now show.
Imagine a sequence along
which $\rho$ is everywhere nonzero
and along which, therefore, $U$ is everywhere defined. Since
each configuration
along the sequence is obtained from preceding configurations
by  continuous transformations, $U$ varies continuously
as $t$ changes, and
$N_H(t)$ is constant.
To obtain $N_H(t_1)\neq N_H(t_2)$, it is
necessary that at some $t$ between $t_1$ and $t_2$
there is a point in
space at which $\rho$ vanishes ($U$ is undefined)
and accordingly the energy density is at
least $\lambda v^4/4$.
While such a configuration does not have arbitrarily
low energy density, it can have arbitrarily low energy.
Consider a configuration with $A_i({\bf x})=0$
everywhere in space, and with
$\rho$ vanishing at one point ${\bf x}_0$.
Further, suppose that $\Phi$ deviates from its
vacuum value only in some region of characteristic
size $L$ about ${\bf x}_0$.  Then,
the energy
of this configuration can be reduced
to an arbitrarily
small value by reducing $L$.
Thus, a sequence of configurations with
$N_H(t_1)\neq N_H(t_2)$ can have arbitrarily
low maximum energy.

In the remainder of this section we show that
for solutions which dissipate at early and late times,
as opposed to more general sequences
of configurations, obtaining a different $N_H$
for $t\rightarrow\infty$ than for $t\rightarrow -\infty$
{\it does} require at least the sphaleron energy.
To this end,
we introduce the gradient
descent integer $N_{GD}$
which characterizes nonvacuum configurations
and which, as explained below,
has the property that $N_{GD}$
changes only if the sphaleron barrier is
crossed\cite{mantonprivate}.

To define $N_{GD}$ of a configuration,
we use that configuration as the $\tau=0$ initial condition
in the equations
\begin{mathletters}
\label{genGDeqns}
\begin{eqnarray}
\partial_\tau A_i
&=& - \frac{\delta V[A_i,\Phi]}{\delta A_i}
\label{AgenGDeqns} \\
\partial_\tau \Phi
&=& - 2\,\frac{\delta V[A_i,\Phi]}{\delta\Phi^*} \ .
\label{PhigenGDeqns}
\end{eqnarray}
\end{mathletters}
These equations evolve the initial $(\tau=0)$
configuration in a direction in configuration
space which is along the gradient of $V$ with the sign
chosen so that $V$ decreases as $\tau$ increases.
(Equations (\ref{genGDeqns}) are the equations
of motion (\ref{eqsofmotion}) with $A_0=0$ and
with second order time
derivatives replaced
by first order $\tau$ derivatives.)
We will assume that there are
no local energy minima in configuration
space except the true vacua with $V=0$.
Then, there are two possible outcomes as $\tau\rightarrow\infty$.
For most configurations,
the gradient
descent equations (\ref{genGDeqns}) evolve the configuration towards
a vacuum configuration as $\tau\rightarrow\infty$.
For such configurations, we define
$N_{GD}$ as the Higgs winding number of
the vacuum configuration reached from the
original configuration as $\tau\rightarrow\infty$.
($N_{GD}$ is gauge invariant under small
gauge transformations, but it changes under large
gauge transformations.)
For vacuum configurations, $N_{GD}=N_H$,
while for nonvacuum configurations $N_{GD}$
is the winding number of the
nearest vacuum, found by sliding
down the potential.
$N_H$ can change during the
slide, and therefore
$N_{GD}$ and $N_H$ can differ.
There are special configurations
for which the gradient descent equations do not
lead to a vacuum configuration as $\tau\rightarrow\infty$.
These configurations mark the boundaries
between the basins of attraction of different
vacua.
For these
configurations, $N_{GD}$ is not defined.

A continuous sequence of configurations parameterized
by $t$
(not necessarily a solution)
which has been put into the $A_0=0$ gauge
by a $t$ dependent gauge transformation
can be characterized by $N_{GD}(t)$, the
gradient descent integer of the configuration
at time $t$.
Consider a sequence for which the gauge invariant integer
$N_{GD}(t_2)-N_{GD}(t_1)$ is nonzero.
At some intermediate time there must
be a configuration in the sequence with
at least the sphaleron energy.
This can be seen by
constructing the following vacuum to vacuum
sequence of configurations:
append to the sequence from $t_1$ to $t_2$
the two sequences of configurations obtained
during the descents from the $t_1$ and
$t_2$ configurations to their respective
vacua.
This vacuum to
vacuum sequence of configurations connects
vacua with different winding numbers.
Therefore, the reasoning of Ref. \cite{manton} can be applied, and
we conclude that for some $t$ between
$t_1$ and $t_2$, the configuration has energy equal
to or greater than the sphaleron energy.
Thus, changing $N_{GD}$ requires
at least the sphaleron energy.

Let us consider a
solution to the Minkowski space
classical equations of motion of the kind
we discussed in Sections 2 and 3.
Since $N_{GD}(t)$ is integer-valued and
is typically not constant, there
will be times $t$ when it is not defined.
However, at very early and late times, when the
solution has all fields approaching
vacuum values,
$N_{GD}(t)$ becomes constant in time
and we can define
\begin{equation}
  \Delta N_{GD} = \lim_{t\rightarrow\infty}N_{GD}(t) -
  \lim_{t\rightarrow -\infty}N_{GD}(t) \ ,
  \label{DeltaNGDdef}
\end{equation}
which is gauge invariant even under large gauge
transformations.
We now show that for such
a solution $\Delta N_{GD}=\Delta N_H$.
This is equivalent to showing that for
sufficiently early and sufficiently late times,
when the gradient descent procedure is performed
on the configuration, the Higgs winding number
does not change during the descent.

First, we gauge transform the solution to $A_0=0$
gauge.  We use
the solution to give the $\tau=0$
initial configurations
in the gradient descent equations
\begin{eqnarray}
\label{GDeqns}
\partial_\tau A_i
&=& D_j F_{ji} + \frac{i g}{4} \left[ \Phi \left( D_i \Phi \right)^\dagger
- \left( D_i \Phi \right) \Phi^\dagger \right]
\nonumber \\
\partial_\tau \Phi
&=& D_i D_i \Phi - \lambda
\left( {\rm Tr}\, \Phi^\dagger \Phi -v^2 \right)\Phi
\end{eqnarray}
obtained from the potential energy (\ref{Veqn}).
As long as $\rho$ does not vanish, it
is convenient to rewrite (\ref{GDeqns})
in terms of $\rho$ and $U$ instead of $\Phi$.
This will be justified {\it a posteriori}
when we
show that for configurations
characteristic of solutions at early and late
times $\rho$ never vanishes during the descent.
It is also convenient to introduce the gauge invariant
variable
\begin{equation}
\label{wdefn}
W_i\, \equiv\, \frac{i}{g}\, U^\dagger D_i U
= U^\dagger A_i U + \frac{i}{g} U^\dagger \partial_i U
= A_i^{U^\dagger} \ .
\end{equation}
Using $\Phi = \rho U$, the definition (\ref{wdefn}),
and with the help of
\begin{eqnarray}
{\rm Tr}\, U^\dagger \partial_\tau U
= &0& \nonumber \\
U^\dagger D_i D_i U = -g^2 W_i W_i &-& i g \partial_i W_i
\label{facts}
\end{eqnarray}
we write the gradient descent equations (\ref{GDeqns})
as
\begin{mathletters}
\begin{equation}
U^\dagger \left(\partial_\tau A_i \right) U
= D^W_j F^W_{ji} -\frac{g^2}{2}\rho^2\, W_i
\label{GDeqn2a}
\end{equation}
\begin{equation}
U^\dagger  \partial_\tau  U
= - ig\,\rho^{-2} \partial_i \left(\rho^2 W_i\right)
\label{GDeqn2c}
\end{equation}
\begin{equation}
 \partial_\tau \rho
= \partial_i \partial_i \rho
-\frac{g^2}{2} \rho \,{\rm Tr}\,(W_i W_i)
- 2\lambda \left(\rho^2 - \frac{v^2}{2}\right)\rho
\label{GDeqn2b}
\end{equation}
\end{mathletters}
where $D^W_j F^W_{ji}$ is defined as
$D_j F_{ji}$ of equations (\ref{DmuFmunu}) and (\ref{Fmunu}) with $A_i$
replaced by $W_i$.  (This is a slight abuse of notation
since $W_i$ is gauge invariant.)
We now use (\ref{GDeqn2a}) and (\ref{GDeqn2c})
to obtain
\begin{equation}
 \partial_\tau  W_i
= D^W_j F^W_{ji} -\frac{g^2}{2}\rho^2 W_i
-ig\,\left[\ W_i \  {\bf ,} \ \rho^{-2}\partial_j
\left(\rho^2 W_j \right)\  \right] + \partial_i
\left\{\rho^{-2} \partial_j \left(\rho^2 W_j \right)
\right\} \ .
\label{GDeqnW}
\end{equation}
We now have gradient descent equations
(\ref{GDeqn2b}) and (\ref{GDeqnW}) involving only
the gauge invariant variables
$W_i$ and $\rho$.  We can solve them
first, and then use (\ref{GDeqn2c}) and (\ref{wdefn})
to obtain $U$ and $A_i$.

We wish to apply the gradient descent procedure
to configurations taken from solutions at
very early and late times when
the energy density is everywhere small.
Small energy density means that $W_i$
and $h=\rho - v/\sqrt{2}$ are everywhere small.
Therefore, we linearize equations (\ref{GDeqn2b})
and (\ref{GDeqnW}) in $W_i$ and $h$, and obtain
\begin{mathletters}
\begin{eqnarray}
 \partial_\tau  W_i &=&
\left(\partial_j \partial_j  - m_W^2\right) W_i
\label{linWeqn}
\\
 \partial_\tau  h &=&
\left(\partial_j \partial_j  -  m_h^2\right) h \ .
\label{hlineqn}
\end{eqnarray}
\end{mathletters}
We start at $\tau=0$ with a configuration
in which $|h|$ is everywhere less than some
$h_0$.
The solution to (\ref{hlineqn})
valid for all $\tau\ge 0$  is determined
by $h$ at
$\tau = 0$ and is
\begin{equation}
h(\tau,{\bf x}) = \frac{{\rm e}^{-m^2 \tau}}{(4\pi\tau)^{3/2}}
\int d^3 {\bf y}\,
{\rm e}^{-({\bf x}-{\bf y})^2/4\tau}\, h(0,{\bf y}) \ .
\label{hsolution}
\end{equation}
Using (\ref{hsolution}) and noting that
$|h(0,{\bf y})|\le h_0$, we find that
\begin{eqnarray}
|h(\tau,{\bf x})|
&\le & \frac{{\rm e}^{-m^2 \tau}}{(4\pi\tau)^{3/2}}\int d^3 {\bf y}\,
{\rm e}^{-({\bf x}-{\bf y})^2/4\tau} h_0 \nonumber \\
&=& {\rm e}^{-m^2 \tau} h_0 \nonumber\\
&\le & h_0 \ .
\label{jeffrey}
\end{eqnarray}
By starting with a configuration taken from
the solution at arbitrarily early or late time, we
can make $h_0$ arbitrarily small and in particular much
less than $v/\sqrt{2}$.
Because
$|h|\le h_0$ for all ${\bf x}$ and for all $\tau \ge 0$,
we conclude that the solution to (\ref{hlineqn})
is arbitrarily close to the solution to the
full nonlinear descent equations.
Therefore, there are no places where $h=-v/\sqrt{2}$
and $\rho=0$ during the descent.  We see that
$N_H$ cannot change during the descent.

We have shown that when configurations with
arbitrarily small energy density are used
as initial conditions for the gradient
descent procedure, $N_H=N_{GD}$.
This implies
that for solutions to the
classical equations of motion
which dissipate at early and late times,
$\Delta N_{GD}= \Delta N_{H}$,
and thus
solutions with $\Delta N_H \neq 0$
have at least the sphaleron energy.\footnote{Our proof
that $\Delta N_H = \Delta N_{GD}$ goes through
for any continuous sequence of configurations
parameterized by $t$ ranging from $-\infty$ to $\infty$
for which the energy density of the configurations
dissipates uniformly in ${\bf x}$ as $t\rightarrow\pm\infty$.
This class of sequences includes the solutions
to the equations of motion we consider in
this paper, but is more general. }

Let us recapitulate.  For vacuum to
vacuum sequences of configurations,
$Q=\Delta N_H=\Delta N_{GD}$ is an integer
which is nonzero only
for sequences which cross
the sphaleron barrier.
For sequences which do not begin
and end in vacuum, the situation
is more complicated.  Such sequences
include solutions, and for solutions
$Q$ cannot be defined.
A nonzero value for
$\Delta N_{GD}$
arises only for sequences which
begin and end near different vacua and which therefore
include
configurations with at least the
sphaleron energy.
However, there are sequences of configurations
for which $\Delta N_H\neq 0$ which only include
configurations with arbitrarily
low energy.
For the particular case
of solutions which dissipate at early and
late times,
$\Delta N_H=\Delta N_{GD}$,
and such solutions therefore have $\Delta N_H \neq 0$
only if they cross the sphaleron barrier.

\vskip .3in

\section{Fermion Production in a Classical Background}
\label{fermsection}

In this section we address the question of
fermion production in the background of the classical
solutions which we have described.
We introduce a quantized fermion field $\Psi$,
and as in the standard electroweak theory, we couple
only the left-handed component of the fermion
to the non-Abelian gauge field and introduce
a Yukawa coupling between the fermion
and the Higgs field to give the fermion
a gauge invariant mass.  The action for the fermion is
\begin{equation}
S^{\rm fermion} = \int d^4 x \bar\Psi \left[
i \gamma^\mu D_\mu - \frac{\sqrt{2}m_f}{v}\left( \Phi P_R
+\Phi^\dagger P_L\right) \right] \Psi \ ,
\label{fermaction}
\end{equation}
where $D_\mu = \partial_\mu - i g A_\mu P_L$,
$P_L= \frac{1}{2}(1-\gamma_5)$ and $P_R=\frac{1}{2}(1+\gamma_5)$.
Here $A_\mu=A_\mu^a \sigma^a /2$ so the left-handed
component of $\Psi$ is an $SU(2)$ doublet.
For simplicity, both the up and the down components
of $\Psi$ have the same mass $m_f$.  The
gauge invariant normal ordered fermion current
\begin{equation}
\label{Jdef}
J^\mu =\, : {\bar\Psi} \gamma^\mu \Psi :
\end{equation}
is not conserved, that is,
\begin{equation}
\label{anomaly}
\partial_\mu J^\mu = \frac{g^2}{32 \pi^2}\, \epsilon^{\mu\nu\alpha\beta}
\,{\rm Tr} \, \left( F_{\mu\nu} F_{\alpha\beta} \right) \ .
\end{equation}

As discussed in the Introduction, in a background
which does not begin and end in pure gauge, fermion production
cannot simply depend on $Q$.  In this section, we
show that in the background of
a solution whose energy density dissipates
for $t\rightarrow\pm\infty$, the number of fermions produced
is $\Delta N_H$.
Our argument applies equally to any background whose
energy dissipates uniformly in ${\bf x}$
at early and late times, solution
or not.
Such backgrounds may
have well-defined
$Q$, integer or non-integer, or they may
have undefined $Q$.
Viewed as an index theorem, our $3+1$ dimensional
non-Abelian result generalizes an index
theorem
of Weinberg \cite{weinberg}
for the 2 dimensional Abelian Higgs model.

To begin, consider as a background not a solution but rather
a sequence of configurations $A_i({\bf x},t)$ and $\Phi({\bf x},t)$
(with $A_0=0$) for $-T\le t\le T$ with the fields pure
gauge at $\pm T$, that is
\begin{mathletters}
\label{puregauge}
\begin{equation}
A_i({\bf x},-T) = \frac{i}{g} U_p({\bf x})\partial_i U_p^\dagger
({\bf x})  ~~~~~~  \Phi({\bf x},-T) = \frac{v}{\sqrt{2}}
U_p({\bf x})
\label{pastpuregauge}
\end{equation}
and
\begin{equation}
A_i({\bf x},T) = \frac{i}{g} U_f({\bf x})\partial_i U_f^\dagger
({\bf x})  ~~~~~~  \Phi({\bf x},T) = \frac{v}{\sqrt{2}}
U_f({\bf x})\ .
\label{futurepuregauge}
\end{equation}
\end{mathletters}
This means that at $t=\pm T$ the fermion Hamiltonian is
gauge equivalent to the free Hamiltonian for a
fermion doublet of mass $m_f$.  Also, assume that the
background is localized in the sense that there is an
$R_T$ such that if $|{\bf x}|>R_T$ then
$A_i({\bf x},t)=0$ and $\Phi({\bf x},t)=v/\sqrt{2}$.
In this case the topological charge $Q$ given
by (\ref{topQ}) with the integral over $t$
going from $-T$ to $T$ is an integer
\begin{equation}
\label{vactovacQ}
Q=w[U_f]-w[U_p]
\end{equation}
which is also the change in Higgs winding $\Delta N_H$
between $-T$ and $T$.  Under these circumstances a
fermion state which at $t=-T$ has $n$ fermions
will evolve into a state with $n+Q=n+\Delta N_H$
fermions at $t=T$.  This result is
established by a direct application of the work
of N. Christ \cite{christ} who studied the evolution of
the in-vacuum, $|0^{\rm in}\rangle$, in
a background which begins and ends in pure gauge.
He showed that $|0^{\rm in}\rangle$, which in the far
past contains no particles, in the far future
is a superposition of states each of which has $Q$ more
fermions than anti-fermions.  Although Ref.
\cite{christ} is restricted to massless fermions, the
formalism is immediately generalizable to massive
fermions as long as the mass terms are gauge
invariant as they are in (\ref{fermaction}).

The field theory calculation of fermion production
is consistent with the more intuitive level crossing
picture.  Consider the instantaneous Hamiltonian
\begin{equation}
\label{fermHam}
{\cal H}(t) = \gamma^0\left[ -i \gamma^i D_i + \frac{\sqrt{2}}{v}\,m_f\,
\left(\Phi P_R + \Phi^\dagger P_L\right)\right]
\end{equation}
which acts on single particle spinor wave functions.
For $|{\bf x}|>R_T$ we have that $A_i({\bf x},t)=0$ and
$\Phi({\bf x},t)=v/\sqrt{2}$
so we can impose periodic spatial boundary conditions on the
wave functions.  This makes the spectrum of ${\cal H}(t)$ discrete.
(We further make the spatial box so large that the level spacings
are much less than $m_f$.)
The spectral flow ${\cal F}$ of ${\cal H}(t)$ is defined
as the number of eigenvalues of ${\cal H}(t)$
which cross zero from below minus the number which
cross zero from above as $t$ ranges from $-T$ to $T$.
Given the conditions (\ref{puregauge})
we know that the spectrum of ${\cal H}(-T)$, which is the
same as the spectrum of ${\cal H}(T)$, has a gap
between $-m_f$ and $m_f$.  Thus ${\cal F}$ can
also be viewed as the number of levels which go from
below $-m_f$ to above $m_f$ minus the number which go from above
$m_f$ to below $-m_f$ as $t$ ranges from $-T$ to $T$.
Now the Atiyah-Patodi-Singer theorem \cite{APS}
(including the identification of the index with
the spectral flow) tells us for the case
at hand that ${\cal F}=Q$.  Thus, the Fock space
calculation which gives that the number of particles
produced is $Q=\Delta N_H$ is consistent with
the intuitive notion that the number of particles
produced is equal to the net number of levels
which cross the mass gap.

We now turn to a background which is a
solution to the Minkowski space equations of motion
which dissipates at early and late times.
In this case, $Q$ is not defined.
However, we are only interested in solutions which
approach pure gauge as $t\rightarrow\pm\infty$.
Thus at very early and very late times the
fermion Hamiltonian is close to the free Hamiltonian
and we should be able to make sense of the question
of how many fermions are produced.

At early and late times $\rho$,
given by
$\Phi({\bf x},t)=\rho({\bf x},t)U({\bf x},t)$,
approaches $v/\sqrt{2}$ and since $\rho$ does not
vanish $U$ can be defined.  At early and late
times the gauge invariant field $W_i=A_i^{U^\dagger}$
of (\ref{wdefn}) also approaches zero.
More specifically, we can choose $T_0$ big
enough to ensure that
\begin{mathletters}
\label{smallness}
\begin{eqnarray}
\Bigl\vert W_i^a({\bf x},-T_0)\Bigr\vert &<& \epsilon\, m_f/4g \nonumber\\
\left\vert\rho({\bf x},-T_0) - v/\sqrt{2}\, \right\vert &<&\epsilon\,
v/8\sqrt{2}
\label{pastsmallness}
\end{eqnarray}
\begin{eqnarray}
\Bigl\vert W_i^a({\bf x},T_0)\Bigr\vert &<& \epsilon\, m_f/4g \nonumber\\
\left\vert \rho({\bf x},T_0) - v/\sqrt{2} \,\right\vert &<&\epsilon\,
v/8\sqrt{2}
\label{futuresmallness}
\end{eqnarray}
\end{mathletters}
where $\epsilon$ is a dimensionless number
which can be made arbitrarily small by going to large enough
$T_0$ and where the constants multiplying $\epsilon$
have been chosen for later convenience.

Now we pick $T_0$ so large that conditions (\ref{smallness})
are satisfied with $\epsilon\ll 1$
and we ask how many fermions
are produced between $-T_0$ and $T_0$.  The formalism
of Ref. \cite{christ} requires that the background
start and end in pure gauge.
We therefore {\it construct} a background
which agrees with our solution for $-T_0\le t\le T_0$
but is pure gauge at $t=\pm T$
where $T>T_0$.
We want the pure gauge configurations
reached at $t=\pm T$
to be close to the configurations of the solution at
$t=\pm T_0$.
By (\ref{smallness}), $A_i({\bf x},\pm T_0) \approx
\frac{i}{g} U({\bf x},\pm T_0)
\partial_i U^\dagger ({\bf x},\pm T_0)$
and $\Phi ({\bf x},\pm T_0)\approx \frac{v}{\sqrt{2}}
U({\bf x},\pm T_0)$, so we choose
the pure gauge configurations
of (\ref{puregauge}) to have
$U_p({\bf x})=U({\bf x},-T_0)$
and $U_f({\bf x})=U({\bf x},T_0)$.
The background we construct is
\begin{mathletters}
\label{interp}
\begin{eqnarray}
\bar\Phi({\bf x},t)&=&\cases{
\displaystyle{\left[ \frac{v}{\sqrt{2}} f(t) + \rho({\bf x},-T_0)
\left(1-f(t)\right)\right]U({\bf x},-T_0)} &
$\ \ \ \ \ \ \ \ \ \ \ \ \ \,-T\le t\le -T_0$ \cr
{}~&~\cr
\rho({\bf x},t)U({\bf x},t) &
$\ \ \ \ \ \ \ \ \ \ \ \ \,-T_0\le t \le T_0$ \cr
{}~&~\cr
\displaystyle{\left[ \rho({\bf x},T_0)\bar f(t) +
\frac{v}{\sqrt{2}} \left(1-\bar f(t)\right)\right]
U({\bf x},T_0)}  &
$\ \ \ \ \ \ \ \ \ \ \ \ \ \ \ T_0\le t\le T$ \cr
}
\label{Phiinterp}\\
{}~&&\nonumber\\
\bar A_i({\bf x},t)&=&\cases{
\displaystyle{
\frac{i}{g} U({\bf x},-T_0)\partial_i U^\dagger({\bf x},-T_0) f(t)
+ A_i({\bf x},-T_0)\left(1-f(t)\right)}  & $\,-T\le t \le -T_0$
\cr ~&~ \cr
A_i({\bf x},t) & $-T_0 \le t\le T_0$ \cr ~&~\cr
\displaystyle{A_i({\bf x},T_0)\bar f(t) +
\frac{i}{g} U({\bf x},T_0)\partial_i U^\dagger({\bf x},T_0)
\left(1-\bar f(t)\right)} & $\ \ \,T_0\le t \le T$ \cr
}
\label{Ainterp}
\end{eqnarray}
\end{mathletters}
where $f(t)$ goes smoothly and monotonically from $1$ to $0$ as $t$
goes from $-T$ to $-T_0$ and $\bar f(t)$ goes smoothly and monotonically
from $1$ to $0$ as $t$ goes from $T_0$ to $T$.
For this background the topological charge
is $Q=w[U({\bf x},T_0)]-w[U({\bf x},-T_0)]$ which is the change
in Higgs winding of the solution between $-T_0$ and $T_0$.
(Recall that $T_0$ has been chosen so large that
$U({\bf x},t)$ does not change its winding for
$|t|>T_0$.)  The number of fermions produced in
the background (\ref{interp}) is therefore
$\Delta N_H$ of the solution.

The background (\ref{interp}) only matches the
solution for $-T_0\le t \le T_0$.  We now argue that
in the background (\ref{interp}) no fermions are
produced while $t$ is between $-T$ and $-T_0$ and
while $t$ is between $T_0$ and $T$.  For $T_0 \le t \le T$
let us examine the spectrum of the single particle
Hamiltonian ${\cal H}(t)$ given by (\ref{fermHam})
with the fields $\bar A_i$ and $\bar \Phi$ of (\ref{interp}).
Since the spectrum is gauge invariant we
can just as well use the fields gauge transformed
by $U^\dagger({\bf x},T_0)$, which for $T_0\le t\le T$
gives
\begin{mathletters}
\label{uninterp}
\begin{equation}
\bar A_i^{U^\dagger ({\bf x},T_0)}({\bf x},t) =
\bar f(t) W_i({\bf x},T_0)
\label{Auninterp}
\end{equation}
\begin{equation}
\label{Phiuninterp}
U^\dagger({\bf x},T_0)\bar \Phi({\bf x},t) = \bar f(t)
\rho({\bf x},T_0) + \frac{v}{\sqrt{2}}\,\left(1-\bar f(t)\right)\ .
\end{equation}
\end{mathletters}
The Hamiltonian (\ref{fermHam}) for $T_0\le t\le T$
now takes the form
\begin{mathletters}
\label{lateHam}
\begin{equation}
{\cal H}(t) = {\cal H}_{\rm free} + {\cal H}'(t)
\label{splitHam}
\end{equation}
where
\begin{equation}
\label{Hprime}
{\cal H}'(t) = \bar f(t) \left[ -g\, \gamma^0 \gamma ^i P_L\,
W_i({\bf x},T_0) + \frac{\sqrt{2}}{v} \, m_f \,\gamma^0
\left( \rho({\bf x},T_0)- \frac{v}{\sqrt{2}}\right) \right] \ .
\end{equation}
\end{mathletters}
For any $n\times n$ Hermitian matrix $M$, the
maximum of the absolute value
of its eigenvalues, $\Vert M\Vert$, is less
than or equal to $n$ times the maximum of the
modulus of its entries.  Now ${\cal H}'(t)$ acts on eight
component spinors and since  ${\cal H}'(t)$ is
already diagonal in the ${\bf x}$ basis we have that
\begin{equation}
\label{eigenvalues}
\Vert {\cal H}'(t) \Vert \leq 8\, \bar f(t) \,
\max_{\bf x} \left\{ \,\frac{g}{2}\, \Bigl\vert W_i^a({\bf x},T_0)\Bigr\vert
\ , \ \frac{\sqrt{2}}{v}\, m_f\, \Bigl\vert \rho({\bf x},T_0)
-v/\sqrt{2}\Bigr\vert\, \right\}\ .
\end{equation}
Because of (\ref{futuresmallness}) we have that
\begin{equation}
\label{smallHprime}
\Vert {\cal H}'(t) \Vert <  \bar f(t)\, \epsilon \, m_f
\le \epsilon \, m_f \ .
\end{equation}
We can make $\epsilon$ arbitrarily small by choosing $T_0$ sufficiently
large.  Equation (\ref{smallHprime})
implies that for any normalized state $|\psi\rangle$
\begin{equation}
\langle\psi| {\cal H}'(t)^2 |\psi\rangle <
\epsilon^2 \, m_f^2
\label{Hprimenorm}
\end{equation}
whereas for ${\cal H}_{\rm free}$ we have that
\begin{equation}
\langle\psi| {\cal H}_{\rm free}^2 |\psi\rangle \ge  m_f^2 \ .
\label{Hfreenorm}
\end{equation}
Let $|\psi\rangle$ be an eigenstate of ${\cal H}(t)$ with
eigenvalue $E$, that is $({\cal H}_{\rm free}+{\cal H}')
|\psi\rangle = E|\psi \rangle$.  Now if $E=0$
we have ${\cal H}_{\rm free}|\psi\rangle = - {\cal H}'|\psi\rangle$
which by (\ref{Hprimenorm}) and (\ref{Hfreenorm})
is impossible.  A simple generalization of this argument
shows that if $E$ is an eigenvalue of ${\cal H}(t)$
for any $t$ between $T_0$ and $T$
it must satisfy
\begin{equation}
\label{Econdition}
|E| > m_f - \epsilon\, m_f \ .
\end{equation}
Thus the spectrum of ${\cal H}(t)$ has no eigenvalues
between $-m_f(1-\epsilon )$ and $m_f(1-\epsilon )$ for all $t$
between $T_0$ and $T$ and we conclude that
no levels can cross zero as $t$ goes from $T_0$ to $T$.

We have not developed a field theoretical
formalism for discussing particle production in a
background which does not go from pure gauge to pure
gauge. For the case at hand, however, with the background
(\ref{interp}) we are confident that no fermions
are produced between $T_0$ and $T$ because
the instantaneous single particle Hamiltonian maintains
a gap from $-m_f(1-\epsilon )$ to $m_f(1-\epsilon )$ between
$T_0$ and
$T$. An identical argument leads to the conclusion
that no fermions are produced between $-T$ and $-T_0$.
Since the number of fermions produced between $-T$ and
$T$ is $\Delta N_H$, we have that the number of fermions
produced between $-T_0$ and $T_0$ is
$\Delta N_H$.  Between $-T_0$ and $T_0$ the
fields in (\ref{interp}) are those of our Minkowski
space solution.  Therefore we conclude that for
$T_0$ arbitrarily large the number of fermions produced
in the background of a solution between $-T_0$ and $T_0$
is $\Delta N_H$.   The fact that the topological
charge is not defined does not alter this conclusion.

Our
result is similar
in spirit to that proposed for massless fermions
coupled to unbroken $SU(2)$ gauge theory
by N. Christ in the concluding
section of Ref. \cite{christ} and considered recently
by Gould and Hsu \cite{GH}.
At early and late times, solutions have the form
(\ref{zeroth}) and (\ref{Afuture}) with
$\omega_k=|{\bf k}|$.  Because we have imposed
the boundary condition (\ref{Abc}),
there is an integer $w[U_f]$
associated with solutions to the equations of motion.
Christ proposed that the number of fermions produced
in a background of this form is $w[U_f]$.
For our solutions in the theory with the
Higgs field, $w[U_f]=\Delta N_H$ and we
have shown that the number of
massive fermions produced
is $\Delta N_H$.  Our result holds even for
fermions with arbitrarily small mass, but
our analysis relies on the existence
of a gap in the fermion spectrum and cannot be
applied to a theory with massless fermions.

We have shown that classical solutions which dissipate
at early and late times must have
at least the sphaleron energy
for the Higgs winding number to change.
This means that for solutions with energies below this threshold,
no fermions
are produced.
As we discussed earlier, it is possible to solve the classical equations of
motion order by order in perturbation theory in $g$ and $\lambda$. This is
equivalent to solving them order by order in the amplitudes of the fields.
Any process that happens only for solutions with energy above some
threshold will never be seen in such a perturbative expansion. Therefore,
fermion number violation does not occur in
backgrounds obtained by solving the
classical equations of motion to any finite order in perturbation
theory.

\vskip .3in

\section{Quantum Scattering}
\label{scattsection}

Until now we have kept
the gauge and Higgs fields classical.  Here, we
make a few remarks on quantum scattering
of massive gauge bosons.  If we were working with
an unbroken $SU(2)$ gauge theory, it would be difficult
to proceed from a
classical treatment to quantum scattering, since the asymptotic
states of the quantum theory are glueballs, which have
no classical counterparts.  However, in the spontaneously
broken theory which we are discussing,
the asymptotic states in the quantum theory are
the massive quanta of the gauge field itself,
whose classical analogues are the solutions
to (\ref{lineqns}).
We do  not attempt a complete
quantum mechanical treatment here.
The point of this section is only to show
that there is a direct quantum analogue of the
result of Section 3 that the topological charge
of solutions cannot be defined.

Consider the theory
whose action is given by (\ref{Action}).
Using standard diagrammatic methods, one can construct the
S-matrix describing the scattering of massive gauge
and Higgs bosons order by order in
perturbation theory.  We do not require
that there be many gauge or Higgs bosons in the initial or
final states.
Because of (\ref{QdiffW}), we
attempt to construct a Chern-Simons
number operator
$\hat N_{\rm CS}$
and then define the topological charge
as the difference between the expectation
value of $\hat N_{\rm CS}$ in the final
and initial states of the scattering process.
Because we are only interested in
$\langle \hat N_{\rm CS}\rangle$ in the initial
and final states which consist of massive gauge bosons propagating
freely, we can use the free field expansion
for the gauge field operator
\begin{equation}
  \label{freefield}
  \hat A_\mu^b(x) = \frac{1}{(2\pi)^{3/2}}\int \frac{d^3 k}
  {\left(2\omega_k\right)^{1/2}} \,
  \sum_\lambda \left[
  e^{-i k \cdot x}
  \epsilon^b_\mu(k,\lambda)
  \hat a^{b}_\lambda(k) +
  e^{i k \cdot x}
  \epsilon^{b*}_\mu(k,\lambda)
  \hat a^{b\dagger}_\lambda(k) \, \right] \ ,
  \end{equation}
where the color index $b$ is not summed over.
The sum on $\lambda$ runs from 1 to 3,
the polarization vectors satisfy
$\epsilon^b_\mu (k,\lambda)k^\mu=0$ and $\epsilon^b_\mu(k,\lambda )
\epsilon^{b\mu}(k,\lambda ')=\delta_{\lambda\lambda '}$,
and the creation and annihilation operators satisfy
$\left[\hat a^b_\lambda(k)\, , \, \hat a^{b'\dagger}_{\lambda'}(k') \right]
= \delta^3(k-k') \delta_{\lambda\lambda'}\delta_{bb'}$.
We now define
\begin{equation}
\label{NCSoperator}
\hat N_{\rm CS} = \frac{g^2}{32\pi^2}\,\epsilon^{\ell m n}
\int d^3x\, {\rm Tr} \,\left[ \hat A_\ell \partial_m \hat A_n
+\frac{2}{3}\,ig \hat A_\ell \hat A_m \hat A_n \right] \ .
\end{equation}
Substituting (\ref{freefield}) in (\ref{NCSoperator}),
normal ordering, and taking either the $t\rightarrow\infty$
or the $t\rightarrow -\infty$ limit,
we obtain
\begin{equation}
\label{NCSresult}
\lim_{t\rightarrow\pm\infty} \hat N_{\rm CS}
= \frac{g^2}{32\pi^2}
\int d^3 k\,\frac{|{\bf k}|}{2\omega_k}\,\sum_b \Bigl[
\hat n_L^b({\bf k}) -\hat n_R^b({\bf k}) \Bigr] \ ,
\end{equation}
where $\hat n_L^b({\bf k})$ and $\hat n_L^b({\bf k})$
are the number operators for
left- and right-handed particles with color index
$b$ and momentum $k$.
(For a particle with color label $b$ and with
positive momentum in the
3-direction,
$\hat a^b_L=(\hat a^b_1 + i \hat a^b_2)/\sqrt{2}$ and
$\hat a^b_R=(\hat a^b_1 - i \hat a^b_2)/\sqrt{2}$
and $\hat n^b_L = \hat a_L^{b\dagger} \hat a^b_L$.)
In taking the $t\rightarrow\pm\infty$ limit
of (\ref{NCSoperator}) to obtain
(\ref{NCSresult}),
any terms which do not have the same number
of creation and annihilation operators vanish
by the Riemann-Lebesgue theorem.
At first glance, it seems that (\ref{NCSresult})
can be used to compute the difference between the
expectation value of $\hat N_{CS}$ in the
initial and final states of a scattering process.
This result would be interesting,
because it is certainly possible to have
perturbative (tree-level, in fact)
$2\rightarrow 2$ scattering processes
in which the difference between the number
of left- and right-handed massive gauge
bosons changes.  In the massless theory
$\omega_k=|{\bf k}|$ and (\ref{NCSresult})
is Lorentz invariant.
In the spontaneously broken theory,
however,
(\ref{NCSresult}) is not Lorentz invariant.
It is not possible, therefore, to define
a Lorentz invariant topological charge as
the difference between $\langle \hat N_{\rm CS}\rangle$
in the initial and final states of a scattering
process.

\vskip .3in

\section{Conclusions}
\label{concl}

Classical backgrounds given by solutions
to the equations of motion have undefined topological
charge, but they can nevertheless be characterized
by the integers $\Delta N_H$ and $\Delta N_{GD}$.
For general sequences of configurations,  $\Delta N_{GD}$
is nonzero only if the sphaleron barrier is crossed.  This is
not true in general for $\Delta N_H$, but for solutions to the
equations of motion  which dissipate at early and late times,
$\Delta N_H=\Delta N_{GD}$. Thus, for dissipative solutions,
the Higgs winding number changes only if the energy is greater
than the sphaleron energy. When a massive
chiral fermion is coupled to such a solution, the
number of fermions produced is given by $\Delta N_H$,
and is not related to the topological charge.

In previous work in unbroken $SU(2)$ gauge theory \cite{FKRS},
it was found that the topological charge of classical solutions
can be nonzero at finite order in perturbation theory.
This might  hint that, in nature, fermion
number violating processes could occur at finite order
in perturbation theory.  Now, with the Higgs field included
in the theory, we see how nature
can avoid this outcome.  First, it is impossible
to define the topological charge of classical solutions.
Second, while at first glance it seems that
topological charge could be generated at finite
order in quantum scattering processes, here too
it turns out to be  undefined.  Third, for classical solutions
we have shown that the number of fermions produced
is not given by the topological charge, but by the
change in Higgs winding number.  This means that
fermion  number violation does not occur at finite order
in classical perturbation theory.  Thus, it seems reasonable
to assume that in quantum scattering there is no
fermion number violation at any finite order in perturbation
theory.  This is not a surprising conclusion. What is surprising
is that we have arrived here not by finding $Q=0$,
but by finding that $Q$ is not well-defined and
does not determine the
number of fermions produced.

\vskip .3in

\acknowledgments

We would like to thank S. Coleman and N. Manton for
particularly enlightening discussions.
We would also like to thank N. Christ,
T. Gould, S. Hsu, V. V. Khoze, V. Rubakov and I. Singer
for useful conversations.  K. R.
acknowledges the hospitality of the Aspen Center for Physics,
where part of this work was done.

\end{document}